\documentclass[smallextended]{svjour3}       
\usepackage{latexsym}
\usepackage{graphics}
\usepackage{graphicx}
\usepackage{enumitem}
\usepackage{fix-cm}
\usepackage{bigstrut}
\usepackage{booktabs}
\usepackage{mathptmx}      
\usepackage{url}
\usepackage[colorlinks,citecolor=blue,urlcolor=blue,linkcolor=blue]{hyperref}
\journalname{Gen. Relativ. Gravit.}
\begin{document}
\title{Charged compact stellar model in Finch-Skea spacetime}
\titlerunning{Charged compact stellar model}        
\author{B. S. Ratanpal \and D. M. Pandya \and R. Sharma \and S. Das}
\authorrunning{Ratanpal \em {et al}} 
\institute{B. S. Ratanpal \at
              Department of Applied Mathematics,\\ 
              Faculty of Technology and Engineering,\\ 
              The M. S. University of Baroda,\\
              Vadodara  390 001, Gujarat, India.\\ 
              \email{bharatratanpal@gmail.com}
           \and
           D. M. Pandya \at
              Department of Mathematics and Computer Science,\\ 
              Pandit Deendayal Petroleum University,\\
              Gandhinagar 382 007, Gujarat, India.\\
              \email{dishantpandya777@gmail.com}
           \and
           R. Sharma \at
           Department of Physics,\\ 
           P. D. Women's College,\\ 
           Jalpaiguri 735 101, India.\\
           \email{rsharma@associates.iucaa.in}
           \and
           S. Das \at
           Department of Physics,\\ 
           P. D. Women's College,\\ 
           Jalpaiguri 735 101, India.\\
           \email{dasshyam321@gmail.com}
}
\date{Received: date / Accepted: date} 

\maketitle

\begin{abstract}
In this paper a compact charged stellar model in the Finch and Skea background spacetime [{\it Class. Quantum Gravity} {\bf6}, 467 (1989)] is presented. The model has been developed by assuming a particular charge distribution within the stellar interior. The model is well behaved and can describe a large class of compact stars. Physical features of the model are studied and in particular we show how the presence of charge can have a non-negligible impact on the mass-radius ($M-R$) relationship of such class of stars.

\end{abstract}

\section{Introduction}
\label{intro}

The influence of electromagnetic field on relativistic relativistic stellar objects has been an area of research interest for many decades. The topic has got a renewed attention in the recent past due to the fact that many observed pulsars have been estimated to have huge electric fields. Although astrophysical systems are expected to be globally charge neutral, at certain evolutionary stages, a charged astrophysical object might exist. It is noteworthy that collapse of a star to a point singularity may be avoided by incorporating charge in its composition since the gravitational attraction will then be counter balanced by the Coulomb repulsion. In a binary pulsar system a star may acquire a net charge by accretion from the surrounding medium to form a charged sphere \cite{Treves,Mak}. It has also been pointed out by Usov \cite{Usov} that in ultra-compact stars like `strange stars' composed of $u$, $d$ and $s$ quarks, the electric field at the surface could be as high as $10^{19}~eV/cm$. Physical behavior of such astrophysical systems can be studied by solving the relevant Einstein-Maxwell system. The exterior spacetime of a  static charged spherically symmetric object is uniquely described by the Reissner-Nordstr\"om metric. The interior solution of such a system is, however, not unique and a large class of physically motivated realistic solutions are available in the literature as can be found in a compilation work of Ivanov \cite{Ivanov}.

Physically motivated models of charged fluid/dust spheres have been developed and analyzed by many researchers over the years. Physical behavior and stability of charged dust have been carried out by Papapetrou \cite{Papapetrou47}, Majumdar \cite{Majumdar47}, Bonner \cite{Bonner60,Bonner65}, to name a few. Stettner \cite{Stettner73} has shown that a homogeneous fluid sphere with a considerable amount of net surface charge is more stable than that of an uncharged sphere. Bekenstein \cite{Bekenstein} studied the dynamics and stability of spherical symmetric systems by generalizing the Oppenheimer-Volkoff equations of hydrostatic equilibrium \cite{OppenhiemerV} and obtained a maximum limit on the electric field that the charged sphere might sustain. Cooperstock and Cruz \cite{Cooperstock78} found a new solution for a uniformly distributed charged dust in equilibrium. Bonner and Wickramasuriya \cite{Bonner75} proposed a charged stellar model showing radially increasing matter density. Analytic solutions to Einstein-Maxwell systems for spherically symmetric objects with a variety of charge distributions  have been developed by Wyman \cite{Wyman}, Adler \cite{Adler}, Kuchowicz \cite{Kuchowicz} and Adams \cite{Adams}. Relativistic compact stars in equilibrium in the presence of charge have been studied by Ray \cite{Ray} and Ghezzi \cite{Ghezzi1,Ghezzi2}. By proposing  specific analytic solutions to charged fluid objects, Whitman and Burch \cite{Whitman81} have discussed stability of such objects. Pant and Sah \cite{Pant79} obtained exact solutions to the Einstein-Maxwell system for a charged fluid distribution which allowed them to regain the Tolman IV solution as a particular case by switching off the charge contribution. Later, Pant and Sah \cite{Pant79} solution was generalized by Tikekar \cite{Tikekar84} who analyzed physical viability of the new class of solutions. By assuming a particular charge intensity profile, Patel and Mehta \cite{Patel95} proposed new exact solutions for charged fluid distributions. Rao and Trivedi \cite{Rao98} developed a formalism to generate solutions to coupled Einstein-Maxwell equations. Maartens and Maharaj \cite{Maartens} analyzed the necessary conditions for a solution of the Einstein-Maxwell system to be regular at the interior of a star. New class of conformally flat solutions was generated by Chang \cite{Chang83} for charged fluid as well as dust distributions. In the recent past, realistic solutions corresponding to Einstein-Maxwell systems describing various physically motivated astrophysical objects, have been developed and analyzed by many researchers which include Ref.~ \cite{Maharaj1,Thiru2,Komathiraj1,Maharaj2,Hansraj,Komathiraj2,Patel87,Singh98,Sharma01,Gupta05,Komathiraj07,Maurya11a,Maurya11b,Maurya11c,Pant12,Maurya15,Murad15,Murad13a,Murad13b,Murad13c,Murad13d,Pant14,Ratanpal15charged,Thomas15charged,Durga,Mak1,Mak2,Lake,Pant1,Pant2}. In the absence of any reliable information about the equation of state (EOS) at extreme densities assumption of one of the metric potentials, based on geometry of the associated $3$-space, has been found to be a reasonable technique to construct both charged and uncharged stellar models. Following this technique, a large class of solutions \cite{Sharma06,Mukherjee01,Mukherjee02,Tikekar90,Tikekar98,Tikekar99,Tikekar05,Thomas05,Thomas07,Paul11,Chattopadhyay10,Chattopadhyay12} for charged as well as uncharged fluid spheres have been developed where one makes use of the Vaidya-Tikekar ansatz \cite{Vaidya82}. Similarly, the Finch-Skea ansatz \cite{Finch89} has also been utilized by many authors (see for example, \cite{Tikekar05,Sharma13,Pandya14}) to develop realistic stellar models. In this work, making use of the Finch-Skea ansatz \cite{Finch89}, we propose a new solution to the Einstein-Maxwell system corresponding to a spherically symmetric charged sphere. The main motivation of this study will be to analyze the influence of electromagnetic field on the $M-R$ relationship a compact star. 

The paper has been organized as follows: In Sec.~\ref{sec2} the field equations governing the spherically symmetric static charged object corresponds to the Einstein-Maxwell system have been laid down. By assuming a particular electric field intensity profile, the field equations have been solved in the Finch-Skea \cite{Finch89} background spacetime. The exterior region of the charged fluid distribution is described by the Riessner-Nordstr\"om metric. The junctions conditions joining the interior and the exterior regions have been obtained in Sec.~\ref{sec3}. In Sec.~\ref{sec4}, by imposing necessary regularity conditions, we have studied physical features of the system and analyzed the impact of electromagnetic field on the gross properties of the star, the $M-R$ relationship in particular. 

\section{\label{sec2}The Einstein-Maxwell system and its solution}

We assume that the interior spacetime of a static spherically symmetric star is described by the metric
\begin{equation}\label{IMetric}
	ds^{2}=e^{\nu(r)}dt^{2}-e^{\lambda(r)}dr^{2}-r^{2}\left(d\theta^{2}+\sin^{2}\theta d\phi^{2} \right).
\end{equation}
The physical behavior of the star can be understood by solving the Einstein-Maxwell field equations 
\begin{equation}\label{FE}
R_{i}^{j}-\frac{1}{2} R \delta_{i}^{j}=8\pi G \left(T_{i}^{j}+E_{i}^{j} \right),
\end{equation}
where, the energy-momentum tensor corresponding to matter and electromagnetic fields are respectively given by 
\begin{equation}\label{Tij}
	T_{i}^{j}=\left(\rho+p \right)u_{i}u^{j}-p\delta_{i}^{j},
\end{equation}
and
\begin{equation}\label{Eij}
	E_{i}^{j}=\frac{1}{4\pi}\left(-F_{ik}F^{jk}+\frac{1}{4}F_{mn}F^{mn}\delta_{i}^{j} \right).
\end{equation}
In Eqs.~(\ref{Tij}) and (\ref{Eij}),  $\rho$ and $p$ denote the energy-density and pressure, respectively and $u^{i}$ is the $4$-velocity of the fluid. The electromagnetic field strength tensor can be derived from the $4$-potential $A_{i}=\left(\phi(r), 0, 0, 0 \right)$ so that
\begin{equation}\label{Fij}
	F_{ij}=\frac{\partial A_{j}}{\partial x_{i}}-\frac{\partial A_{i}}{\partial x_{j}}.
\end{equation}
The EM field strength tensor satisfies the Maxwell equations
\begin{equation}\label{ME1}
	F_{ij,k}+F_{jk,i}+F_{ki,j}=0,
\end{equation}
and
\begin{equation}\label{ME2}
	\frac{\partial}{\partial x^{k}}\left(F^{ik}\sqrt{-g} \right)=4\pi\sqrt{-g}J^{i}.
\end{equation}
The $4$-current vector can be obtained in terms of the charge density $\sigma$ using the relation
\begin{equation}\label{Ji}
	J^{i}=\sigma u^{i}.
\end{equation}
Due to spherical symmetry, the only non-vanishing component of $F_{ij}$ is 
\begin{equation}\label{F01}
	F_{01}= -F_{10} = -\frac{e^{\frac{\nu+\lambda}{2}}}{r^{2}}\int_{0}^{r} 4\pi r^{2}\sigma e^{\lambda/2}dr.
\end{equation}
The total charge inside a radius $r$ is given by 
\begin{equation}\label{qr}
	q(r)=4\pi\int_{0}^{r} \sigma r^{2}e^{\lambda/2}dr.
\end{equation}
Note that $E^{2}=-F_{01}F^{01}$ so that the electric field intensity $E$ is given by
\begin{equation}\label{E}
	E=\frac{q(r)}{r^{2}}.
\end{equation}

The field equations given by (\ref{FE}) for the spacetime described by the metric (\ref{IMetric}) are then equivalent to the following set of three non-linear ODE's
\begin{equation}\label{FE1}
	8\pi G \left(\rho+\frac{E^{2}}{2}\right)=\frac{1}{r^{2}}-e^{-\lambda}\left(\frac{1}{r^{2}}-\frac{\lambda'}{r} \right),
\end{equation} 
\begin{equation}\label{FE2}
	8\pi G \left(p-\frac{E^{2}}{2}\right)=e^{-\lambda}\left(\frac{1}{r^{2}}+\frac{\nu'}{r} \right)-\frac{1}{r^{2}},
\end{equation}
\begin{equation}\label{FE3}
	8\pi G \left(p+\frac{E^{2}}{2}\right)=\frac{e^{-\lambda}}{4}\left[2\nu''+\left(\nu'-\lambda' \right)\left(\nu'+\frac{2}{r} \right) \right],
\end{equation}
where, we have set $c = 1$. Note that a prime ($\prime$) denote differentiation with respect to $r$. We, therefore, have a system of five unknowns, namely $\rho$, $p$, $E$, $\nu$ and $\lambda$.

To solve the system, we use the Finch-Skea \cite{Finch89} ansatz  
\begin{equation}\label{ELambda}
e^{\lambda}=1+\frac{r^{2}}{L^{2}}.
\end{equation}
where $L$ is the curvature parameter having dimension of length. The Finch and Skea \cite{Finch89} ansatz is physically well motivated and has been used by many in the past to construct viable stellar models.
Substituting (\ref{ELambda}) in (\ref{FE1}), we get 
\begin{equation}\label{rho1}
	8\pi G \left(\rho+\frac{E^{2}}{2}\right)=\frac{3+\frac{r^{2}}{L^{2}}}{L^{2}\left(1+\frac{r^{2}}{L^{2}} \right)^{2}}.
\end{equation}
Combining (\ref{FE2}) and (\ref{FE3}) and using (\ref{ELambda}) we get,
\begin{equation}\label{NLEq1}
	-\frac{\frac{r^{2}}{L^{2}}}{L^{2}\left(1+\frac{r^{2}}{L^{2}} \right)^{2}}-\frac{1}{\left(1+\frac{r^{2}}{L^{2}} \right)}\left(\frac{\nu{''}}{2}+\frac{\nu{'}^{2}}{4}-\frac{\nu'}{2r} \right)+\frac{r}{L^{2}\left(1+\frac{r^{2}}{L^{2}} \right)^{2}}\frac{\nu'}{2}+8 \pi G E^{2}=0.
\end{equation}
By making a coordinate transformation $z=\sqrt{1+\frac{r^{2}}{L^{2}}}$ and introducing $X=e^{\nu/2}$, equation (\ref{NLEq1}) takes the form
\begin{equation}\label{LEq1}
	\frac{d^{2}X}{dz^{2}}-\frac{2}{z}\frac{dX}{dz}+\left[\frac{(z^{2}-1)-E^{2}L^{2}z^{4}}{z^{2}-1} \right]X=0.
\end{equation}
To solve Eq.~(\ref{LEq1}), we choose the electric field intensity as
\begin{equation}\label{E2}
	E^{2}=\frac{\alpha^{2}(z^{2}-1)}{L^{2}z^{6}},
\end{equation}
so that Eq.~(\ref{LEq1}) reduces to 
\begin{equation}\label{LEq2}
	z^{2}\frac{d^{2}X}{dz^{2}}-2z\frac{dX}{dz}+\left(z^{2}-\alpha^{2} \right)X=0.
\end{equation}
Our choice (\ref{E2}) ensures that $E=0$ at $r=0$. This particular choice also makes the Eq.~(\ref{LEq2}) tractable and its solutions is obtained as
\begin{equation}\label{F}
	X=z^{3/2}\left[C_{1}J_{\beta}(z)+C_{2}Y_{\beta}(z) \right],
\end{equation}
where $\beta=\frac{\sqrt{9+4\alpha^{2}}}{2}$, $C_{1}$, $C_{2}$ are constants of integration, and $J$ and $Y$ are Bessel's functions of first and second kind, respectively. 

The interior spacetime  of the spherically symmetric charged fluid sphere finally takes the form
\begin{equation}\label{IMetric2}
	ds^{2}=z^{3}\left[C_{1}J_{\beta}(z)+C_{2}Y_{\beta}(z) \right]^{2}dt^{2}-\left(1+\frac{r^{2}}{L^{2}} \right)dr^{2}-r^{2}\left(d\theta^{2}+\sin^{2}\theta d\phi^{2} \right).
\end{equation}
Subsequently, the energy density and pressure are obtained in the form
\begin{equation}\label{rho2}
	8\pi G \rho=\frac{2z^{2}(2+z^{2})-\alpha^{2}(z^{2}-1)}{2L^{2}z^{6}},
\end{equation}

\begin{eqnarray}\label{p}
	8\pi G p&=&\frac{\left[3J_{\beta}(z)-z^{2}J_{\beta}(z)+2z\dot{J}_{\beta}(z) \right]C_{1}+\left[3Y_{\beta}(z)-z^{2}Y_{\beta}(z)+2z\dot{Y}_{\beta}(z) \right]C_{2}}{L^{2}z^{4}\left[C_{1}J_{\beta}(z)+C_{2}Y_{\beta}(z) \right]} \nonumber\\
	&& +\frac{\alpha^{2}\left(z^{2}-1 \right)}{2L^{2}z^{6}},
\end{eqnarray}
where an overhead dot (.) denotes differentiation with respect to $z$.

\section{\label{sec3}Junction Conditions}

The spacetime metric (\ref{IMetric2}) must be matched to the exterior Riessner-Nordstr{\"o}m metric
\begin{equation}\label{EMetric}
	ds^{2}=\left(1-\frac{2M}{r}+\frac{Q^{2}}{r^{2}} \right)dt^{2}-\left(1-\frac{2M}{r}+\frac{Q^{2}}{r^{2}} \right)^{-1}dr^{2}-r^{2}\left(d\theta^{2}+\sin^{2}\theta d\phi^{2} \right),
\end{equation}
at the boundary of the star $r=R$. The radius of the star can be obtained by utilizing the condition $p(r=R)=0$. The above boundary conditions determine the constants as
\begin{equation}\label{C1}
	C_{1}=\frac{6z_{R}^{2}Y_{\beta}(z_{R})-2z_{R}^{4}Y_{\beta}(z_{R})+4z_{R}^{3}\dot{Y}_{\beta}(z_{R})+\alpha^{2}\left(z_{R}^{2}-1 \right)Y_{\beta}(z_{R})}{B},
\end{equation}
\begin{equation}\label{C2}
	C_{2}=-\frac{6z_{R}^{2}J_{\beta}(z_{R})-2z_{R}^{4}J_{\beta}(z_{R})+4z_{R}^{3}\dot{J}_{\beta}(z_{R})+\alpha^{2}\left(z_{R}^{2}-1 \right)J_{\beta}(z_{R})}{B},
\end{equation}
where, 
	\begin{eqnarray}\label{B}
	B &=& z_{R}^{5/2}\left\{J_{\beta}(z_{R})\left[6z_{R}^{2}Y_{\beta}(z_{R})-2z_{R}^{4}Y_{\beta}(z_{R})+4z_{R}^{3}\dot{Y}_{\beta}(z_{R})+\alpha^{2}\left(z_{R}^{2}-1 \right)Y_{\beta}(z_{R}) \right]\right.\nonumber\\
	&& \left.-Y_{\beta}(z_{R})\left[6z_{R}^{2}J_{\beta}(z_{R})-2z_{R}^{4}J_{\beta}(z_{R})+4z_{R}^{3}\dot{J}_{\beta}(z_{R})+\alpha^{2}\left(z_{R}^{2}-1 \right)J_{\beta}(z_{R}) \right] \right\}.
\end{eqnarray}
The total mass $M$ is obtained as
\begin{equation}\label{M}
	M=\frac{L\left(z_{R}^{2}-1 \right)^{3/2}\left[z_{R}^{4}+\alpha^{2}\left(z_{R}^{2}-1 \right) \right]}{2z_{R}^{6}},
\end{equation}
where $z_{R}=\sqrt{1+\frac{R^{2}}{L^{2}}}$.
\section{\label{sec4}Physical Analysis}
A physically viable stellar model should satisfy the following conditions throughout the stellar configuration: \\

\noindent (i) $ \rho \geq 0$, ~~~ $p \geq 0 $; \\

\noindent (ii) $ \rho - 3 p   \geq 0 $;\\

\noindent (iii) $ \frac{d\rho}{dr} < 0 $, ~~~ $\frac{dp}{dr} < 0 $;\\

\noindent (iv) $ 0 \leq \frac{dp}{d\rho} \leq 1 $.\\
  
\begin{figure}[ht]
    \includegraphics[width=0.75\textwidth]{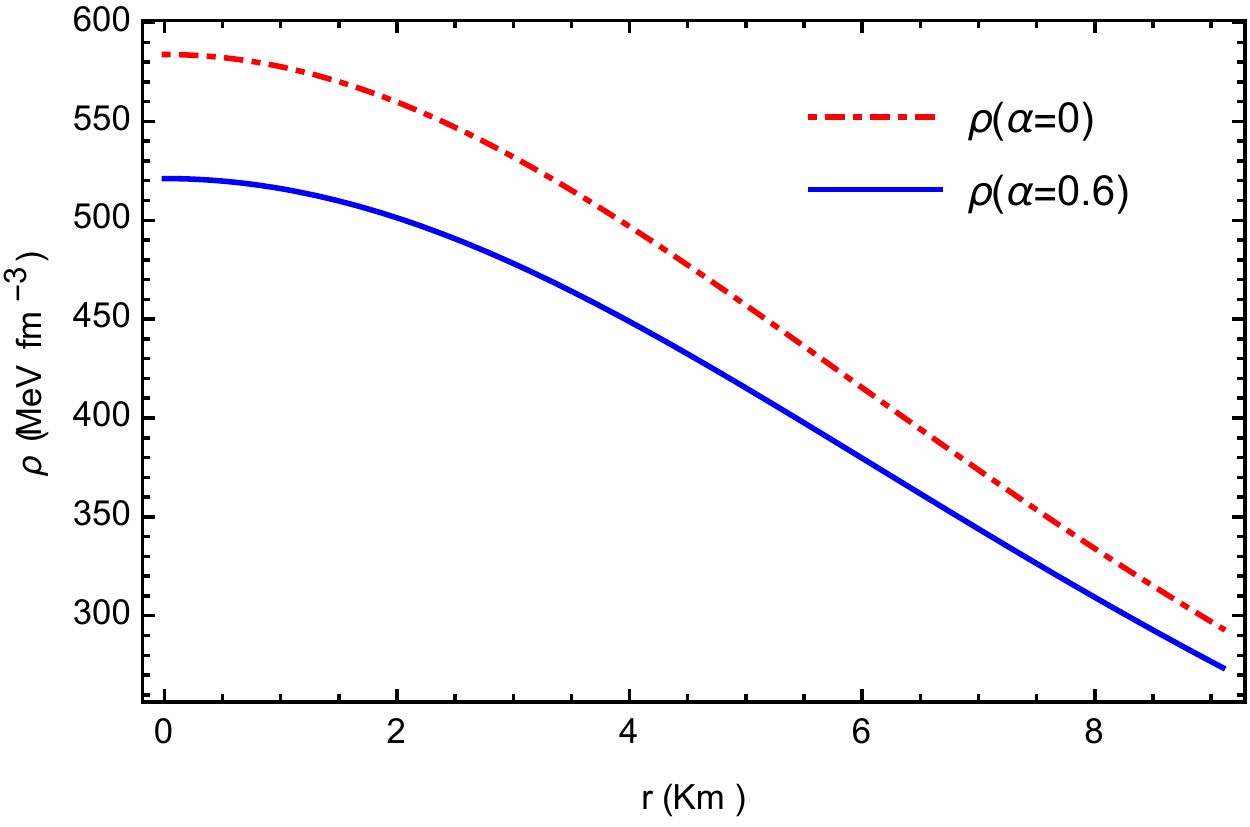}														
\caption{Density profile.}
\label{fig:1}       
\end{figure}

\begin{figure}[ht]
    \includegraphics[width=0.75\textwidth]{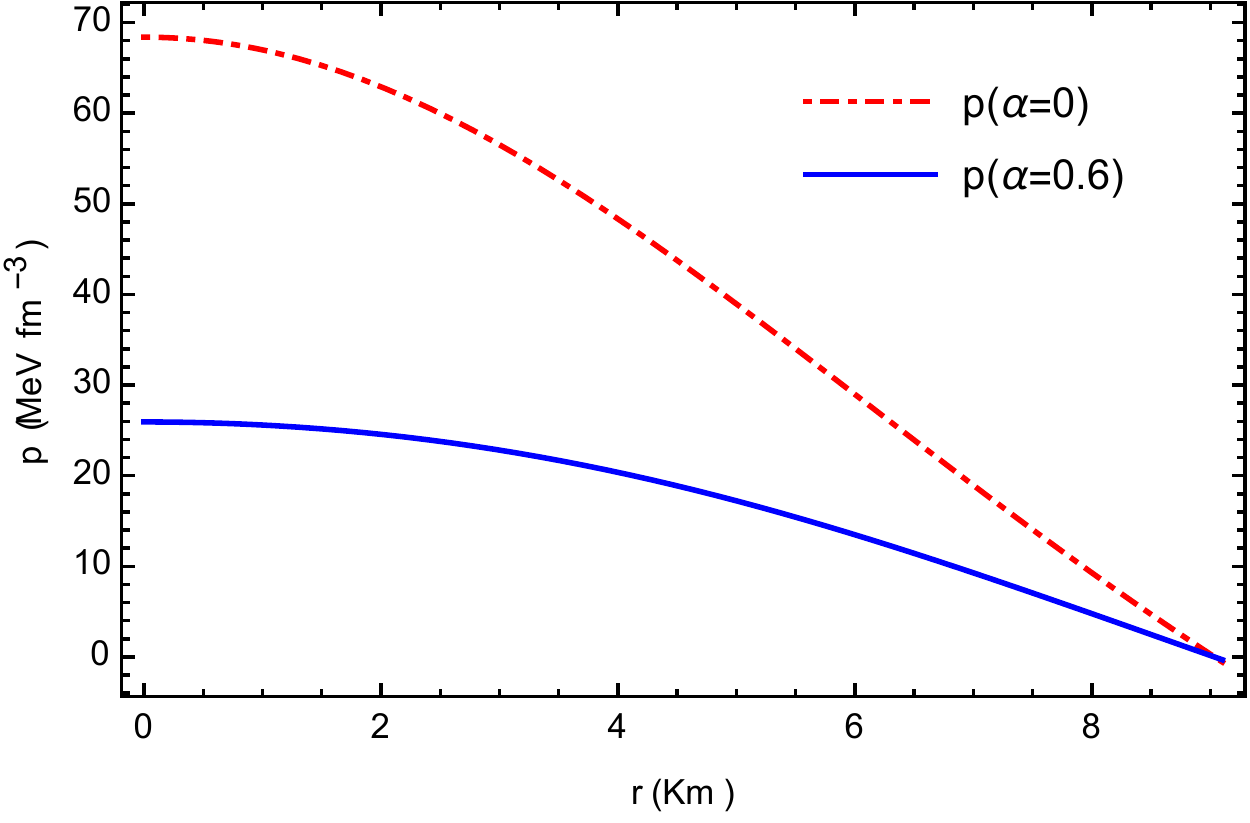}														
\caption{Pressure profile.}
\label{fig:2}       
\end{figure}

\begin{figure}[ht]
    \includegraphics[width=0.75\textwidth]{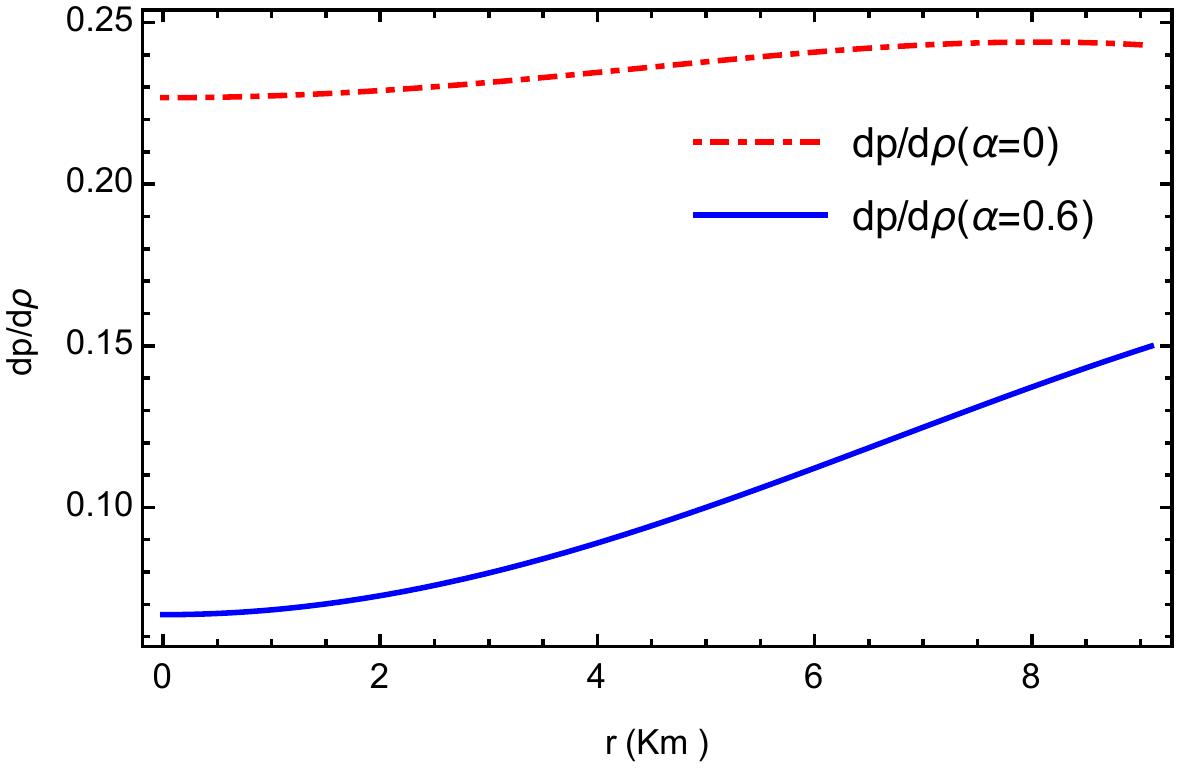}														
\caption{Velocity of sound at the stellar interior.}
\label{fig:3}       
\end{figure}

\begin{figure}[ht]
    \includegraphics[width=0.75\textwidth]{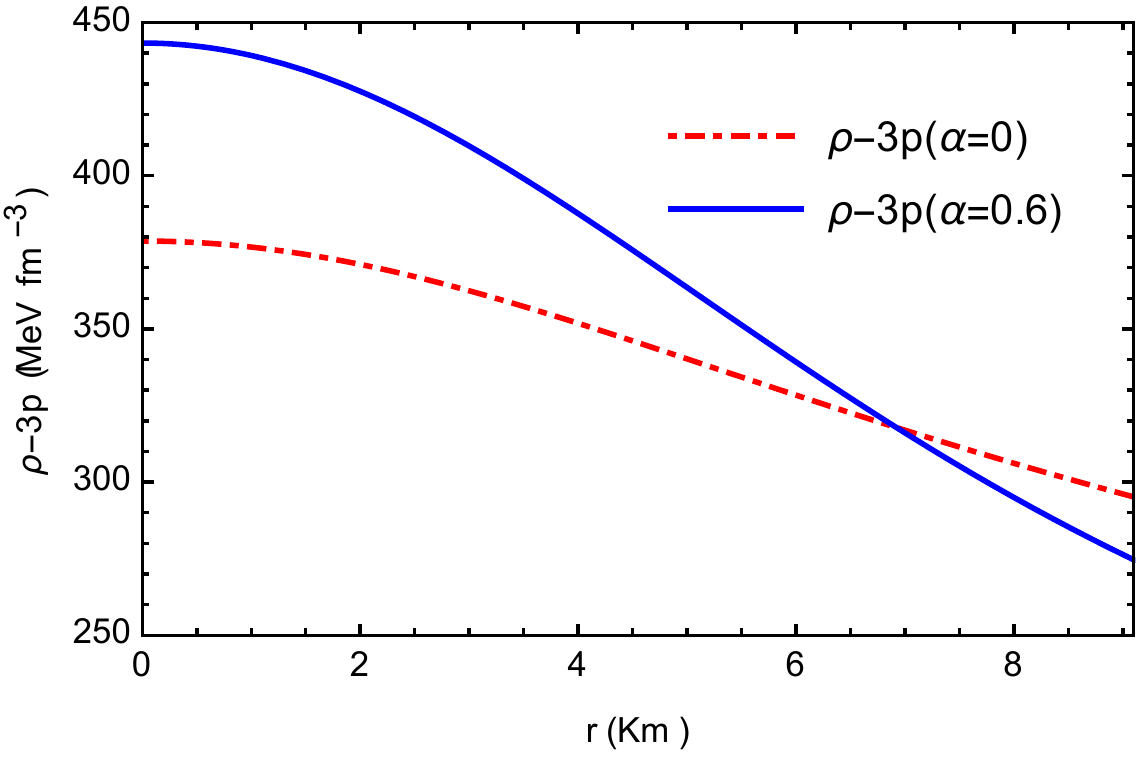}														
\caption{Satisfaction of energy condition within the star.}
\label{fig:4}     		  
\end{figure}

\begin{figure}[ht]
\resizebox{0.75\textwidth}{!}{%
    \includegraphics[width=0.75\textwidth]{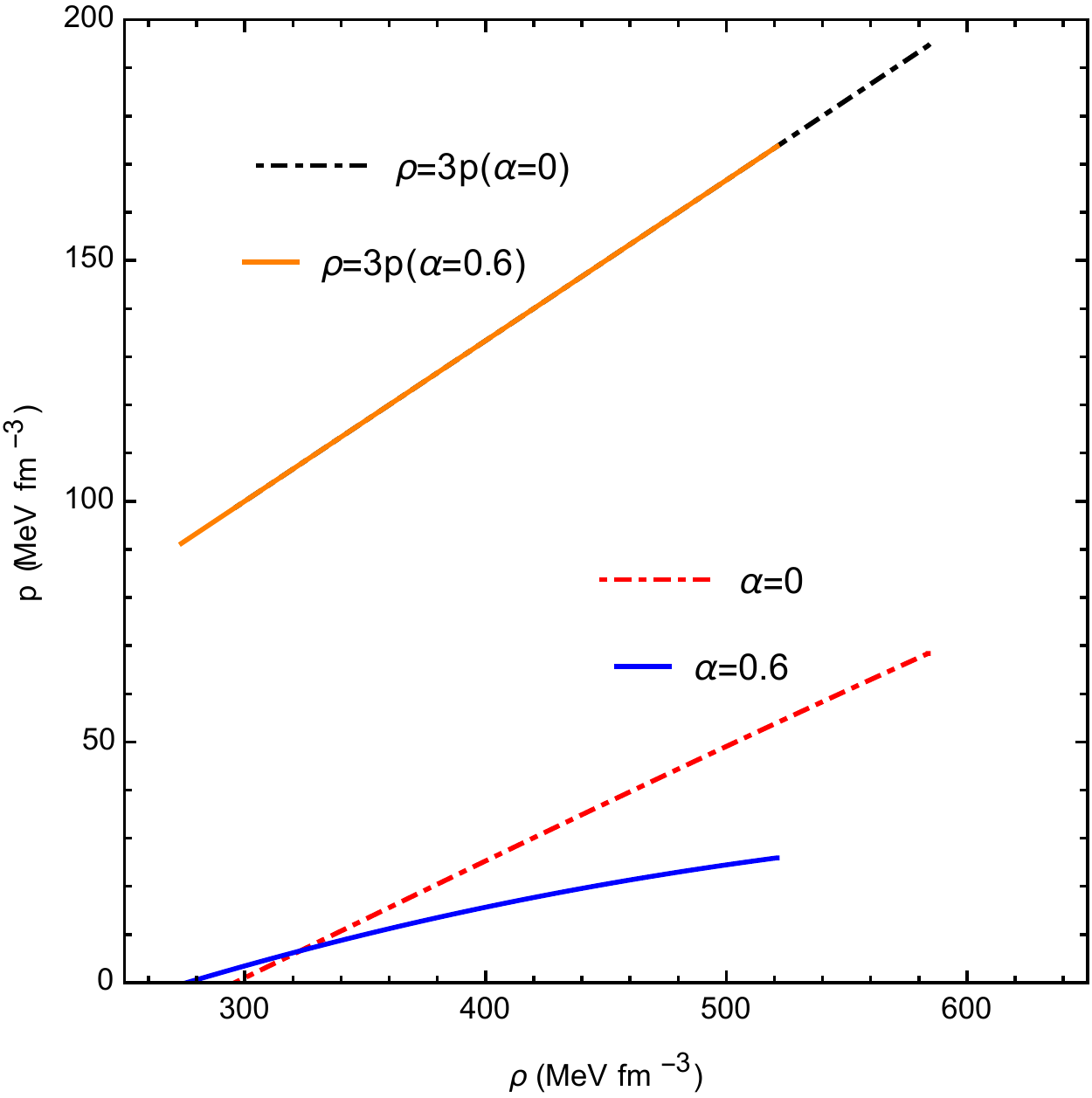}
    }
\caption{EOS profile.}
\label{fig:5}       	
\end{figure}

To examine whether the developed model is regular, well-behaved and capable of describing realistic stars, we have considered the data of the Pulsar $4U 1820-30$ whose mass and radius have recently been estimated to be $M = 1.58~M_\odot $ and $ R = 9.1 $ km, respectively \cite{Gangopadhyay13}. Using these values as inputs in our model, we have determined the values of the other model parameters $C_1$, $C_2$ and $L$. The obtained values ($ C_1 = 0.34$ for $\alpha=0$ and $0.35$ for $\alpha=0.6$; $C_2 = -0.55$ for $\alpha=0$ and $-0.54$ for $\alpha=0.6$) have been used  to show graphically in Fig.~(\ref{fig:1})-(\ref{fig:4}) that all the requirements of a realistic star are satisfied in our model both for charged as well as uncharged cases.  From the figures, it is interesting to note that both density and pressure take lower values if charge is incorporated into the system. Fig.~(\ref{fig:5}) shows thermodynamic behaviour ($p=p(\rho)$) of the system both in charged and uncharged cases. No matter whether charge is incorporated or not, the relationship between radial pressure and density turns out to be linear in this model.

\begin{figure}[ht]
\resizebox{0.75\textwidth}{!}{%
  \includegraphics{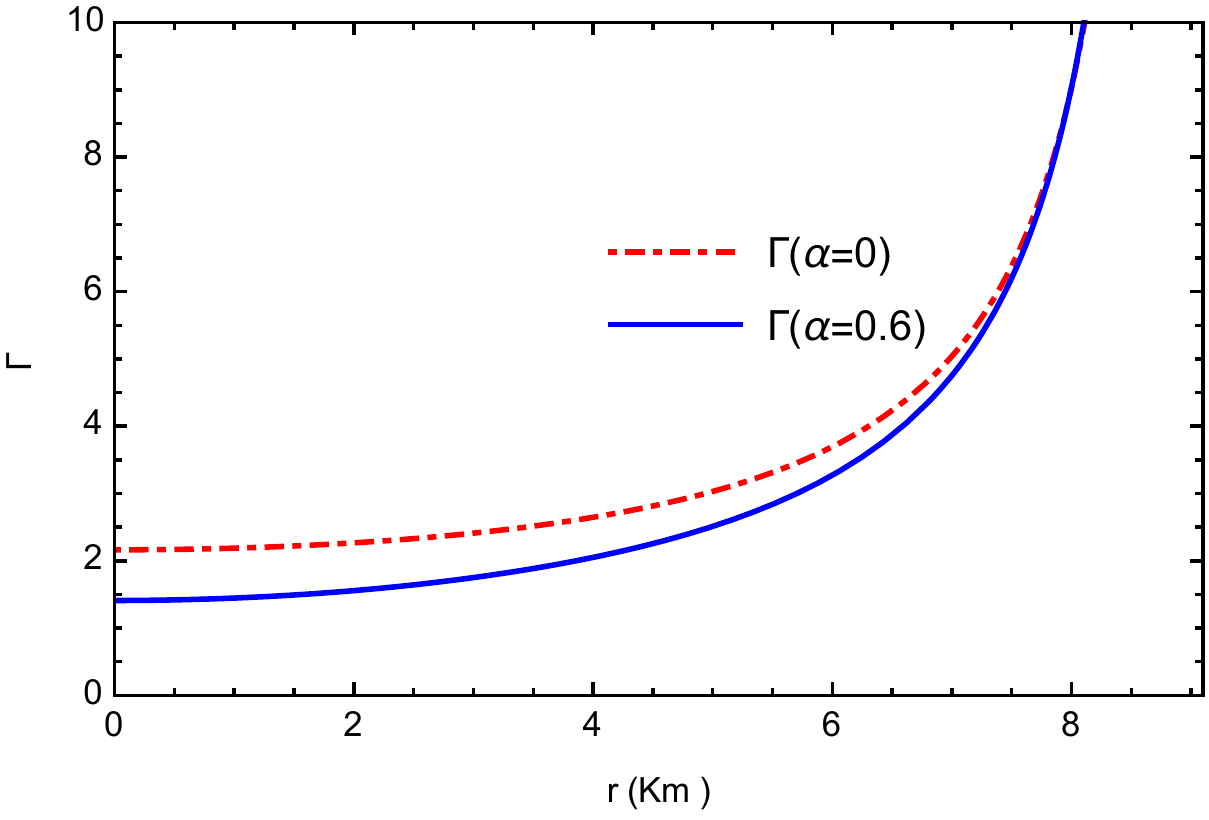}															
}
\caption{Variation of adiabatic index.}
\label{fig:6}       
\end{figure}

We have also calculated the adiabatic index in our model by using the relation 
\begin{equation}
\Gamma=\frac{\rho+p}{p}\frac{dp}{d\rho}
\end{equation} 
Note that for the developed stellar configuration to be stable it is expected that $\Gamma$ should be greater than $4/3$ \cite{hm}. In our model, this condition puts a restriction on $\alpha$ such that $0 \leq \alpha \leq 0.62$. Variation of the adiabatic index for both the charged and uncharged cases has been shown in Fig.~(\ref{fig:6}). We observe a marginal decrease in the values of $\Gamma$ near the center for a charged sphere as compared to its neutral counterpart. 

\begin{figure}[ht]
    \includegraphics[width=0.75\textwidth]{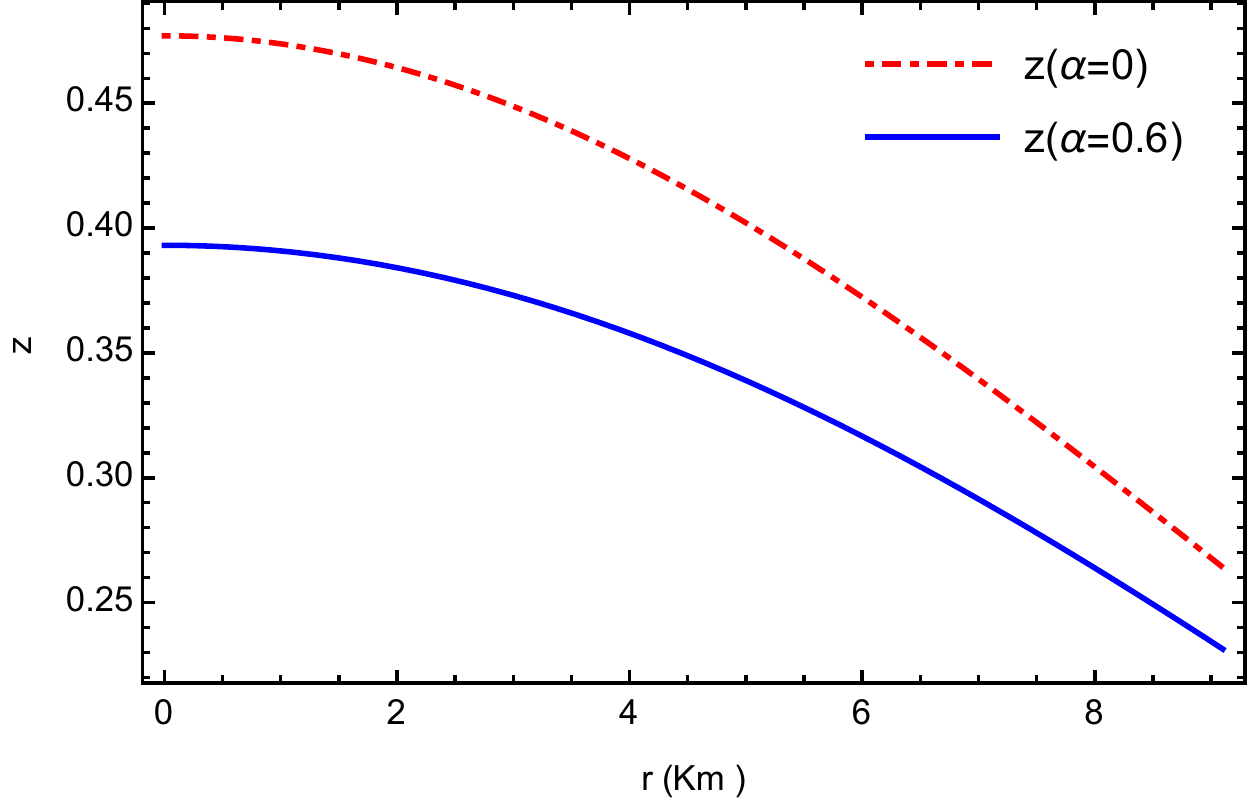}											
\caption{Variation of gravitational redshift.}
\label{fig:7}       	
\end{figure}

\begin{figure}[ht]
    \includegraphics[width=0.75\textwidth]{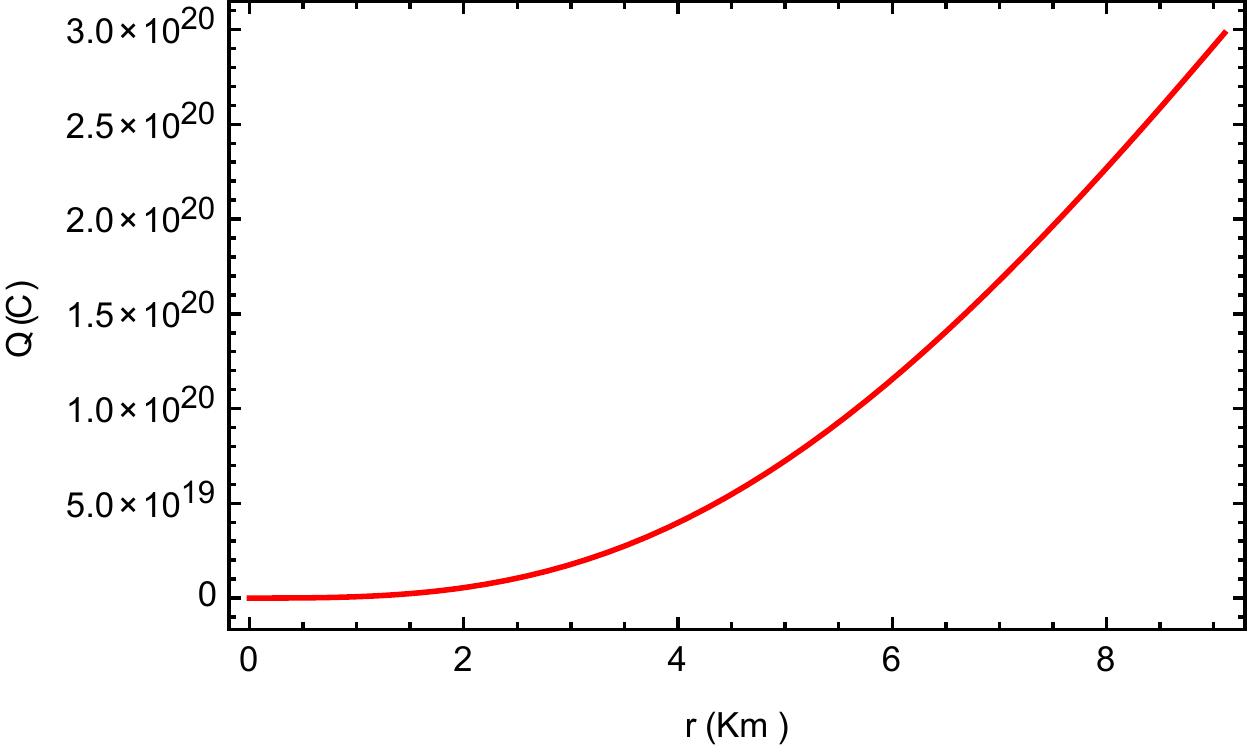}															
\caption{Distribution of charge at the stellar interior.}
\label{fig:8}       
\end{figure}

The surface redshift
\begin{equation} 
z=\left(1-\frac{2 M}{R}\right)^{-\frac{1}{2}}-1,
\end{equation}
also decreases when charge is incorporated into the system as shown in Fig.~(\ref{fig:7}). Radial variation of charge $Q$ has been shown in Fig.~(\ref{fig:8}) which shows that the maximum deposition of charge occurs near the boundary of the star. 

\begin{figure}[ht]
\resizebox{0.75\textwidth}{!}{%
    \includegraphics{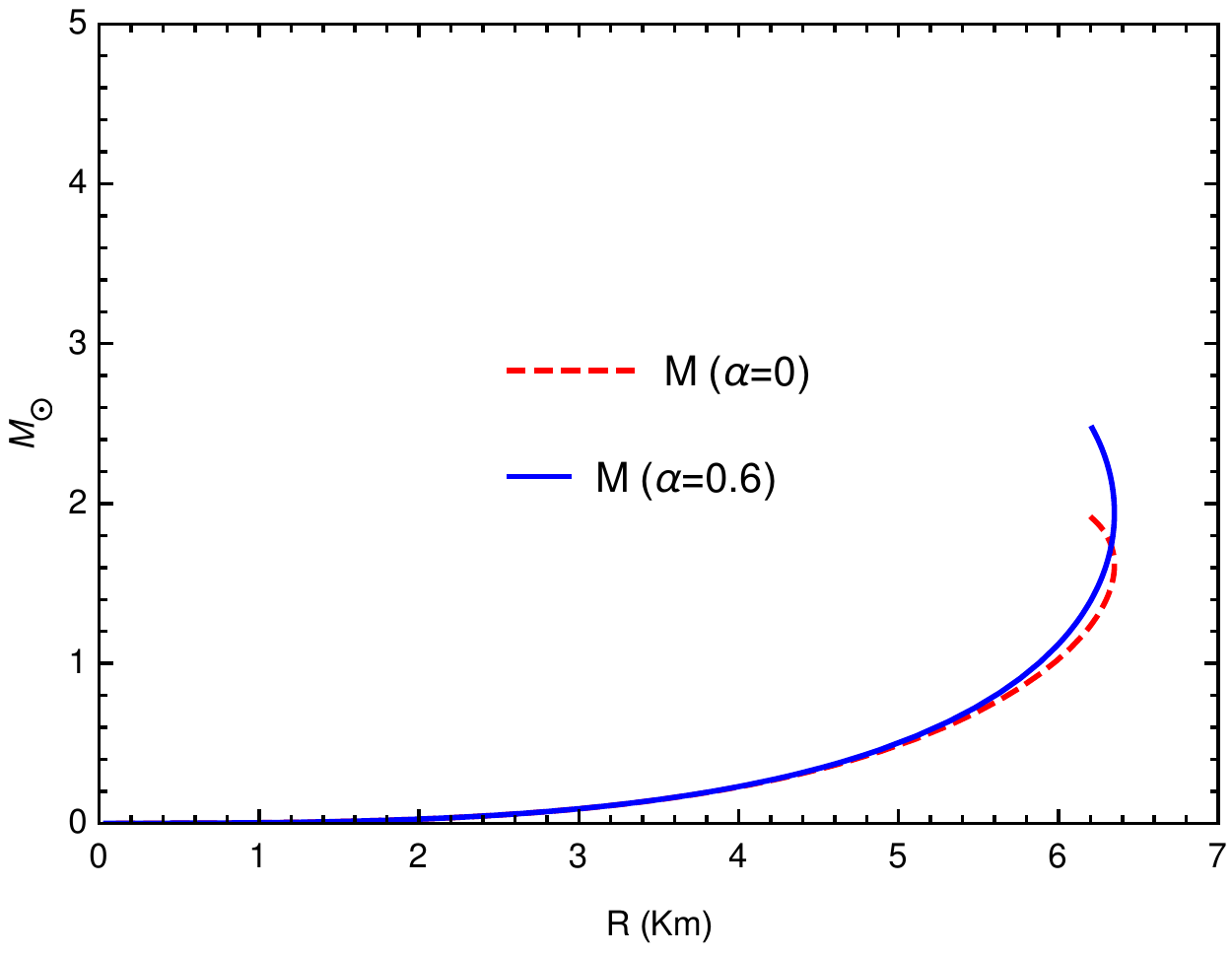}
    }
\caption{Mass-radius relationship. (Assumed surface density $=1.5\times 10^{15}~$gm~cm$^{-3}$.)}
\label{fig:9}       	
\end{figure}

Finally, for an assumed value of the surface density ($\rho(r=R) = 1.5\times 10^{15}~$gm~cm$^{-3}$), we have shown the mass-radius ($M-R$) relationship in this model for $\alpha=0$ and $\alpha \neq 0$ in Fig.~(\ref{fig:9}). This has been done in the following way. For a given surface density, we have fixed the radius $R$ in terms of the curvature parameter $L$ and used the same value of $R$ in the junction conditions to determine the mass $M$. The parametric representation of $R$ and $M$ in terms of $L$ has allowed us to plot the $M-R$ relationship which has been shown in Fig.~(\ref{fig:9}). It is interesting to note that inclusion of charge can accommodate more mass within a stellar configuration. In other words, compactness of a star increases in the presence of charge. 

\section{\label{sec5}Discussion}
By considering the Finch-Skea \cite{Finch89} ansatz and a particular choice of electric field intensity, we have found a non-singular solution of the Einstein-Maxwell System, satisfying all regularity conditions. Stability condition puts a bound on the electric field intensity parameter $\alpha$ = $\left[0, 0.62 \right]$ in this model. One noticeable feature of the model is that in presence of charge, both density $\rho$ and pressure $p$ take lower values as compared to uncharged configurations.  The surface redshift in the charged case appears to be less than that of the uncharged case. In our work, we have analyzed of the mass-radius (M-R) relationship of the model which, in general, is obtained by integrating the TOV equations for a given EOS. In the present study, even though we have not specified any EOS of the matter composition, we have been able to generate the $M-R$ relationship by choosing a particular surface density. We note that, in the presence of charge, stellar configurations can accommodate more mass. It should be stressed here that even though most of the observed pulsar masses are clustered around $1.4~M_{\odot}$, a wide range of values of masses and radii of a large class of pulsars are available for which we still do not have any proper theoretical understanding. Our study shows that in classical gravity, electromagnetic field can be used as a tool to fine-tune the mass vis-a-vis compactness of observed pulsars.

\begin{acknowledgements}
RS gratefully acknowledges support from the Inter-University Centre for Astronomy and Astrophysics (IUCAA), Pune, India, under its Visiting Research Associateship Programme. BSR, DMP and SD are also grateful to the IUCAA for its hospitality where part of this work was carried out.
\end{acknowledgements}

\pagebreak

\end{document}